# Heat and lost work in irreversible cycles: the two sides of the same coin

Joaquim Anacleto

Departamento de Física da Escola de Ciências e Tecnologia da Universidade de Trás-os-Montes e Alto Douro, Quinta de Prados, 5000-801 Vila Real, Portugal

IFIMUP-IN e Departamento de Física e Astronomia, Faculdade de Ciências, Universidade do Porto, R. do Campo Alegre s/n, 4169-007 Porto, Portugal

e-mail: anacleto@utad.pt

ORCID: orcid.org/0000-0002-0299-0146

## Abstract

We discussed the criterion for usefully comparing an irreversible cycle with a reversible one. Grounded on entropy generation, it is proposed a new definition of lost work, which cannot be determined by the usual cycle diagrams, contrary to what has been found in the literature. To better understand the lost work concept and its role in entropy generation, we presented two irreversible cycles, which also instructively reveal that *heat* and *lost work* are the two sides of the same coin, this coin being the *entropy generation*. This study, addressing issues that are virtually absent from the literature, is expected to be not only relevant from a scientific standpoint but also useful for physics teachers and students with a solid background in thermodynamics.

## 1. Introduction

Thermodynamic cycles are important, as they are suitable for introducing some concepts in addition to having useful practical applications. Some of the ones that are usually referred to in textbooks are analyzed in Ref. [1], which covers cycles that exchange heat with only two reservoirs. Among these, the Carnot engine (CE) and the Carnot refrigerator (CR) are particularly relevant because, being reversible, they are the best that is possible for such devices in terms of performance. Interestingly, the Carnot cycle is the unique reversible cycle that operates with only two reservoirs [1].

Although CEs and CRs cannot be realized in practice, they are conceptually very valuable because, besides being a benchmark for real cycles, they can also be used as *auxiliary processes* in some situations to draw general conclusions. In other words, since no entropy is generated by CEs or CRs, its inclusion in a process under analysis results in a process that is identical to the original one [2]. This is what is usually done, for instance, when demonstrating the well-known Clausius theorem [3].

Fig. 1(a) is a diagram of a CE operating between the reservoir temperatures $T_1$ and $T_2$, with $T_1 > T_2$, exchanging heats $Q_1$ and $Q_2$ with the hottest and the coldest reservoir, respectively, and producing the work $W$. Taking energies entering the system as positive and negative otherwise, we have $Q_1 > 0$, $Q_2 < 0$ and $W < 0$.

By reversing the CE operating direction, as illustrated in Fig. 1(b), we obtain a CR for which heats and work keep their absolute values and only change sign, i.e. $Q_1$, $Q_2$ and $W$ in CE become $-Q_1$, $-Q_2$ and $-W$ in CR.

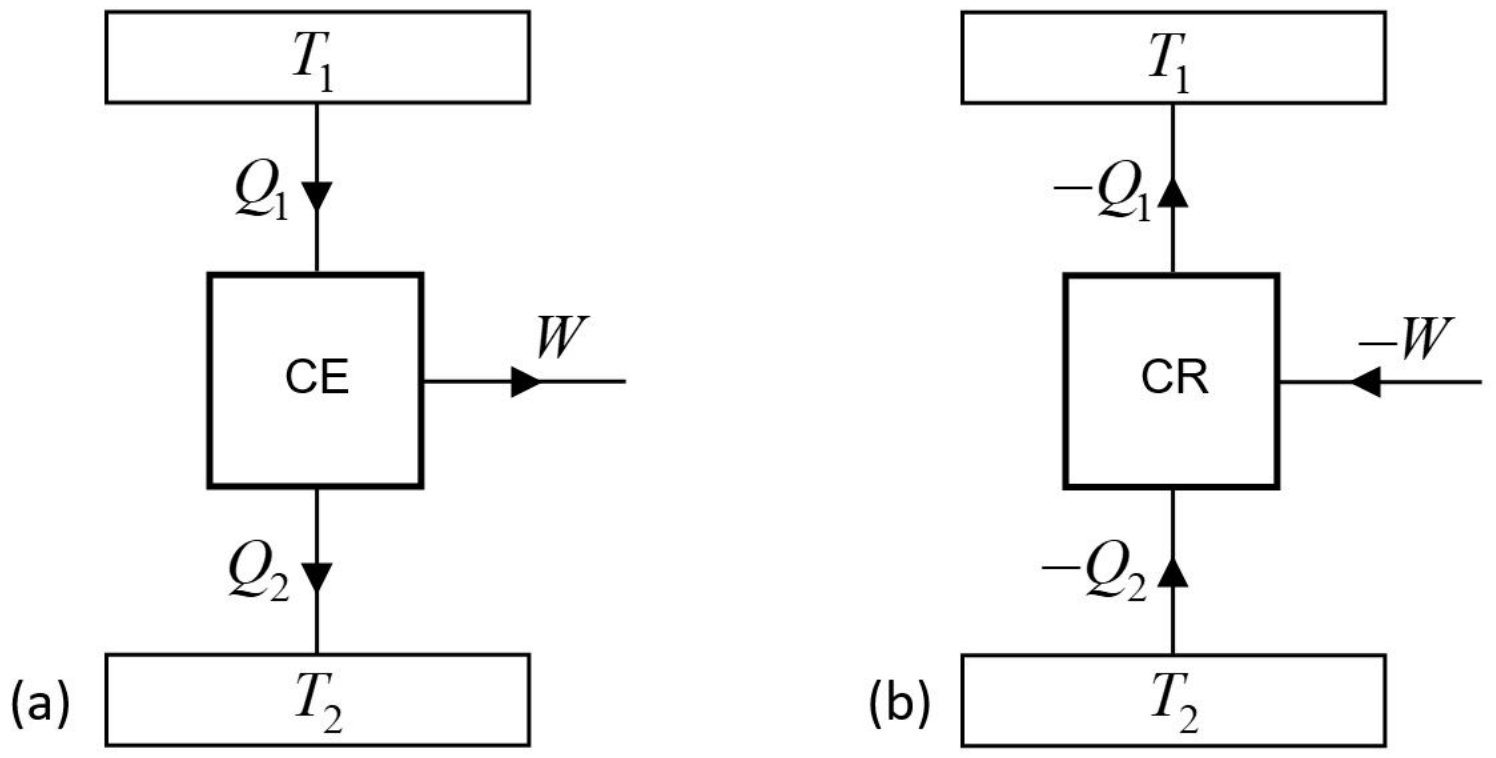

**Figure 1.** (a) Carnot engine (CE) and (b) Carnot refrigerator (CR), both operating between the reservoir temperatures $T_1$ and $T_2$, with $T_1 > T_2$. As said in the text, $Q_1 > 0$, $Q_2 < 0$ and $W < 0$.

The irreversibility of a process (cyclic or not) is gauged by *entropy generation* $\Delta S_{\mathrm{G}}$, which is the sum of system and surroundings entropy variations, i.e.

$$\Delta S_{\mathrm{G}} = \Delta S + \Delta S_{\mathrm{e}}, \qquad (1)$$

with the subscript 'e' denoting surroundings variables. For cyclic processes, we have $\Delta S = 0$ and, by (1), $\Delta S_{\mathrm{G}} = \Delta S_{\mathrm{e}}$. As explained in Ref. [4] (see Appendix), if surroundings consist of reservoirs, the entropy generation $\Delta S_{\mathrm{G}}$ can be given either in terms of *infinitesimal heat* $\delta Q$ and reservoir temperature $T_{\mathrm{e}}$,

$$\oint \frac{-\delta Q}{T_{\mathrm{e}}} = \Delta S_{\mathrm{G}}, \qquad (2)$$

or in terms of *infinitesimal lost work* $\delta W_{\mathrm{L}}$ and system temperature $T$ at which it takes place,

$$\oint \frac{\delta W_{\mathrm{L}}}{T} = \Delta S_{\mathrm{G}}. \qquad (3)$$

Since CE and CR are reversible cycles, $\Delta S_{\mathrm{G}} = 0$ and it follows from (2) that

$$\frac{Q_1}{T_1} + \frac{Q_2}{T_2} = 0, \qquad (4)$$

and from (3), as $\delta W_{\mathrm{L}} \geq 0$ [4], that the lost work $W_{\mathrm{L}}$ is zero,

$$W_{\mathrm{L}} = \oint \delta W_{\mathrm{L}} = 0. \qquad (5)$$

Once the CE and the CR have been described, the purpose is to consider their irreversible counterparts, and so this paper will continue with that task. Section 2 is devoted to discussing the criterion for usefully comparing the *irreversible engine* (IE) with the CE and the *irreversible refrigerator* (IR) with the CR. It is argued that what meaningfully distinguishes an irreversible cycle from its reversible counterpart is the *entropy generation* $\Delta S_{\mathrm{G}}$ and the role of lost work in assessing irreversibility [5-9] is clarified. In section 3, two irreversible cycles exchanging heat with two reservoirs are analyzed, and it is shown that the *heat* exchanged at given *reservoir temperatures* and the *lost work* that takes place at given *system temperatures* are two possible and alternative descriptions of entropy generation $\Delta S_{\mathrm{G}}$. This stems from the formal similarity between (2) and (3). The conclusions are presented in section 4 and, finally, the Appendix, though not strictly necessary, summarizes some ideas taken from Ref. [4] for quick reference.

To our knowledge, the discussion carried out herein complements other studies, e.g. [1, 10], and is not found in other sources. This work develops a new approach to determining entropy generation in irreversible cycles and, besides its scientific character, is intended to be useful for physics teachers and students.

## 2. Irreversible engine (IE) and refrigerator (IR)

Let us now consider the irreversibility of heat engines and refrigerators, which exchange heat with two reservoirs. The CE of Fig. 1(a) is taken as the basis for the discussions throughout this paper, and therefore $Q_1 > 0$, $Q_2 < 0$, $W < 0$, $T_1$, and $T_2$ are reference values, with $T_1 > T_2$.

Fig. 2(a) shows an IE that receives the same energy input $Q_1$ as the CE of Fig. 1(a). As we now have $\Delta S_{\mathrm{G}} > 0$, the heat rejected by this IE to the coldest reservoir is necessarily greater than the one rejected by the CE, which can be stated as $Q_2 - E$, where $E > 0$. By the first law, the irreversible work $W_{\mathrm{I}}$ is smaller than the reversible one delivered by the CE and is given by $W_{\mathrm{I}} = W + E$. Imposing the same input energies for both CE and IE is a possible criterion when comparing the two, which is also the one adopted in Ref. [5].

The IR, i.e. the CR irreversible counterpart, can now be obtained from the previous IE by taking the symmetrical values of $Q_1$, $Q_2$ and $W$, while $E$ is kept the same, as is illustrated in Fig. 2(b). The IE and IR thus obtained lead to the same *entropy generation* $\Delta S_{\mathrm{G}}$, which, by (2) and (4), is related to $E$ by

$$\Delta S_{\mathrm{G}} = \frac{E}{T_2}. \tag{6}$$

By contrast, it is not possible to establish a unique relation between total *lost work* $W_{\mathrm{L}}$ and $E$, since, by using (3) and (6), what we can get is just

$$\oint \frac{\delta W_{\mathrm{L}}}{T} = \frac{E}{T_2}. \tag{7}$$

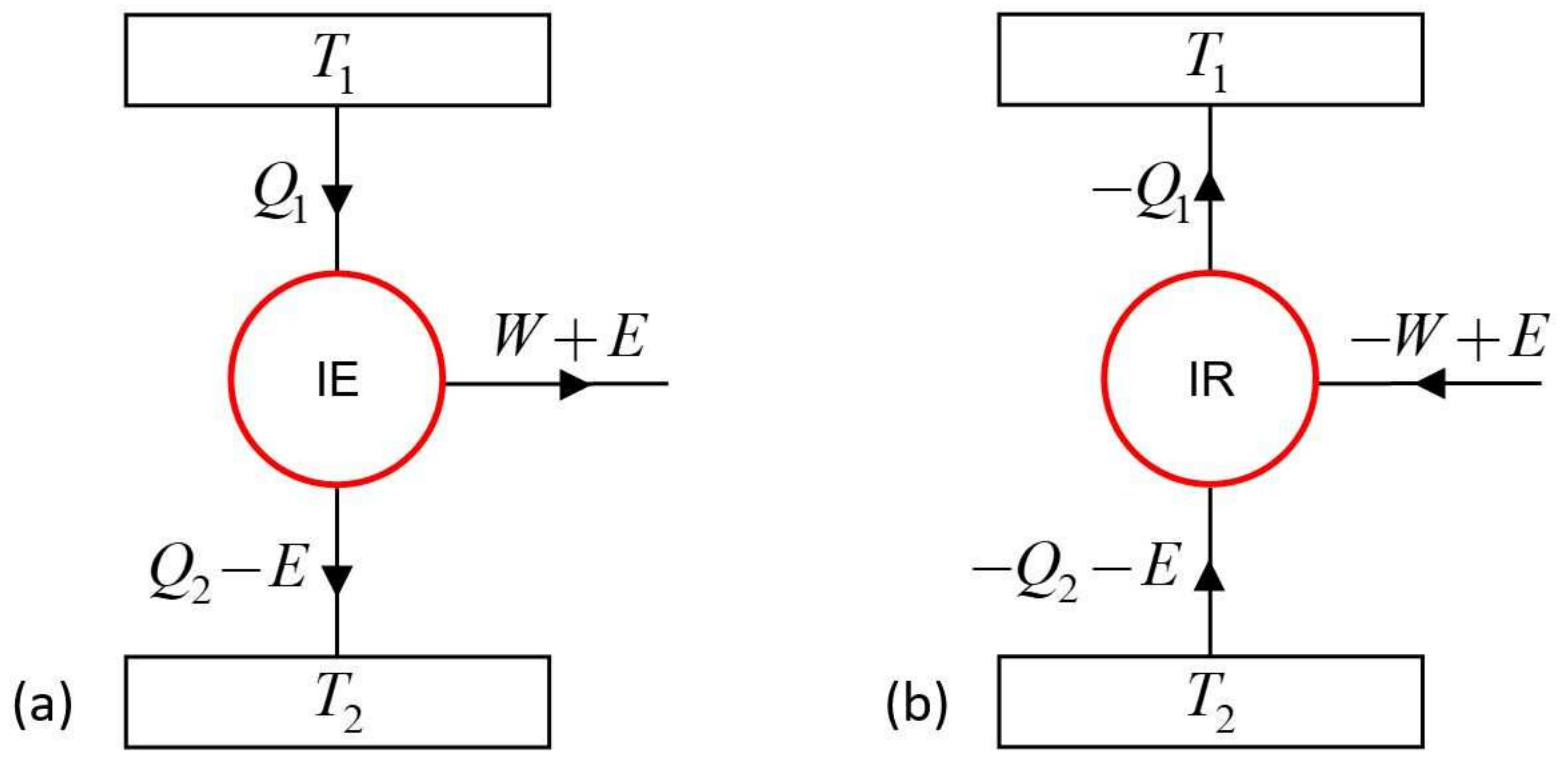

**Figure 2.** (a) IE and (b) IR, both operating between reservoir temperatures $T_1$ and $T_2$, with $T_1 > T_2$. Absolute values of the heats exchanged with the hottest reservoir are equal and the same as that for the CE and CR. $E > 0$ is related to the irreversibility.

Considering Fig. 2, $E$ can be interpreted as either the excess heat rejected by the IE (when compared to that rejected by the CE) or the less heat absorbed by the IR (when compared to that absorbed by the CR). However, (7) does not allow a unique relation between total *lost work* $W_\mathrm{L} = \oint \delta W_\mathrm{L}$ and $E$, since the system temperature $T$ is undefined or unknown. We only have $W_\mathrm{L} = E$ when in (7) $T$ is equal to $T_2$, but that is a particular case.

By looking at Figs. 1 and 2, we see that $E$ can also be interpreted as the difference between irreversible work $W_\mathrm{I}$ and reversible work $W_\mathrm{R}$, i.e.

$$E = W_\mathrm{I} - W_\mathrm{R} \,. \tag{8}$$

Even though $E$ as stated by (8) is in some contexts defined as lost work (e.g. [5]), we claim that lost work $W_\mathrm{L}$ is most properly defined in its differential form by $\delta W_\mathrm{L} = T\mathrm{d}S_\mathrm{G}$ [4]. As we proceed it will become clear that the proposed definition relates to entropy generation in a more fundamental way than $E$ does. For instance, (6) is neither general nor fundamental because it results from the assumption that the heat exchanged with the reservoir at temperature $T_1$ for both the reversible cycle and corresponding irreversible one is the same.

If it makes sense to consider that the heat exchanged with the reservoir at temperature $T_1$ is the same when comparing engines, such a requirement is questionable when comparing refrigerators, because the function of the latter is distinct from that of the former. Thus, we can obtain the IR from the CR alternatively by keeping either the same $-Q_2$ or the same $-W$, as shown in Figs. 3(a) and 3(b), respectively, with $E' > 0$ and $E'' > 0$. It is important to notice that for the IR of Fig. 3(b), $E''$ cannot be identified as $W_\mathrm{I} - W_\mathrm{R}$ because in this case $W_\mathrm{I} = W_\mathrm{R} = -W$.

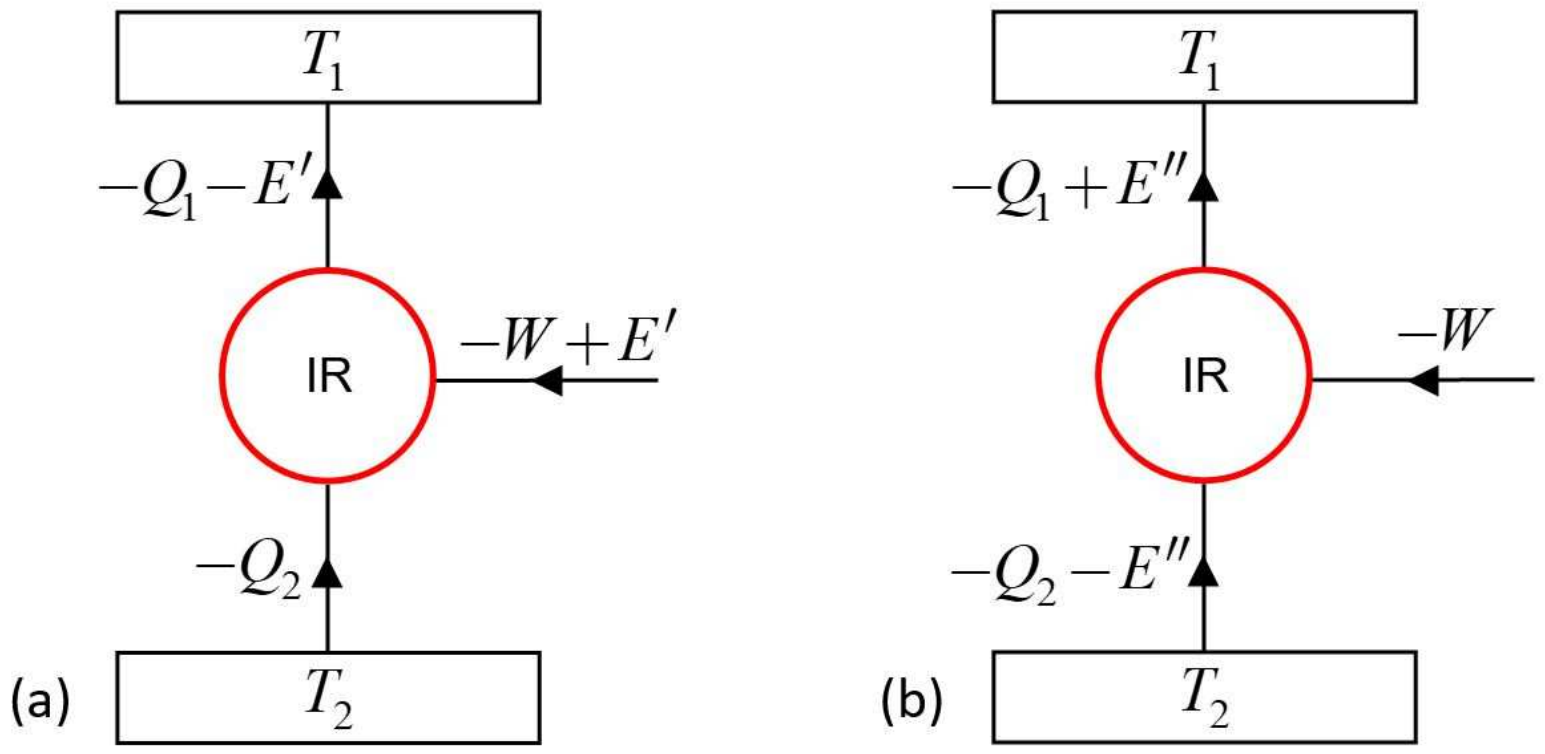

**Figure 3.** Alternatives for setting up the IR from the CR: (a) the heat $-Q_2$ is kept the same; (b) the work $-W$ is kept the same. $E' > 0$ and $E'' > 0$ relate to irreversibility.

Since no relation has been imposed between $E$, $E'$ and $E''$, entropy generations for the IRs of Figs. 2 and 3 are also unrelated and may therefore be different. So, which IR should be chosen as the CR irreversible counterpart? This indetermination is solved by considering the same *entropy generation* $\Delta S_{\rm G}$ for all IRs, i.e. by (2) and (4)

$$\Delta S_{\rm G} = \frac{E}{T_2} = \frac{E'}{T_1} = \frac{E''}{T_2} - \frac{E''}{T_1}, \tag{9}$$

which, according to Ref. [2], is the necessary and sufficient condition for them all to be identical to each other. This means that once a cycle is completed by one given IR, auxiliary Carnot cycles can be used to modify the surroundings in such a way that they become indistinguishable from those obtained by another IR characterized by the same entropy generation.

So, for example, let us show that if (9) holds, the IR of Fig. 3(a) is identical to that of Fig. 2(b). Consider the former together with an auxiliary CE adjusted so that it rejects the heat $-E$ to the coldest reservoir, as shown in Fig. 4. For this auxiliary CE, the absorbed heat from the hottest reservoir is $ET_1/T_2$ and the delivered work is $-\eta ET_1/T_2$, where $\eta$ is the Carnot efficiency given by

$$\eta = 1 - \frac{T_2}{T_1}. \tag{10}$$

Provided (9) holds, $E' = ET_1/T_2$ and it immediately follows that the combined IR of Fig. 4 is *indistinguishable* from that of Fig. 2(b). In other words, since the auxiliary CE does not add extra irreversibility, the IR of Fig. 3(a) has the same irreversibility as that of Fig. 2(b).

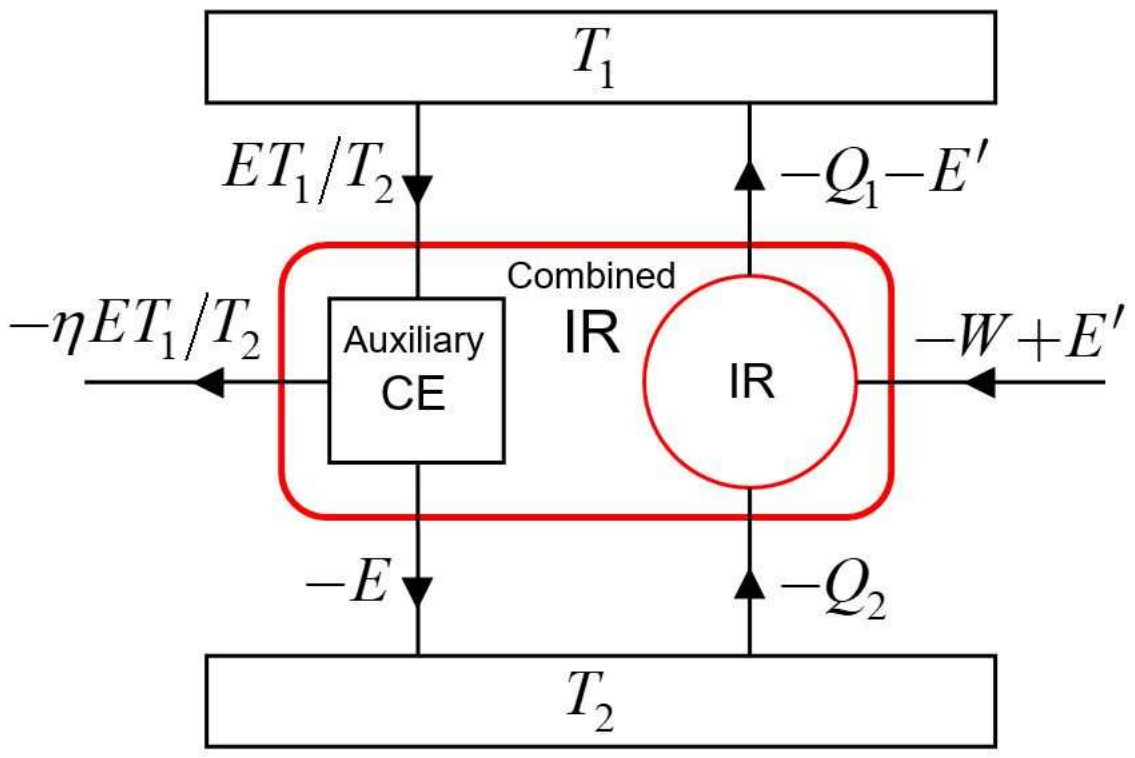

**Figure 4.** IR of Fig. 3(a) together with an auxiliary CE adjusted to reject the heat $-E$ to the coldest reservoir. As no extra irreversibility is added by the CE, by imposing (9), the combined IR thus obtained is *indistinguishable* from that of Fig. 2(b).

Using Fig. 5, parallel reasoning can be done to show that, if (9) holds, the IR of Fig. 3(b) is identical to that of Fig. 2(b). Moreover, other similar reasoning easily convinces us that *all* IRs *with the same entropy generation* $\Delta S_{\rm G}$ are identical to each other, so either one can be used to analyze irreversibility. This statement also applies to IEs and is in line with Ref. [2].

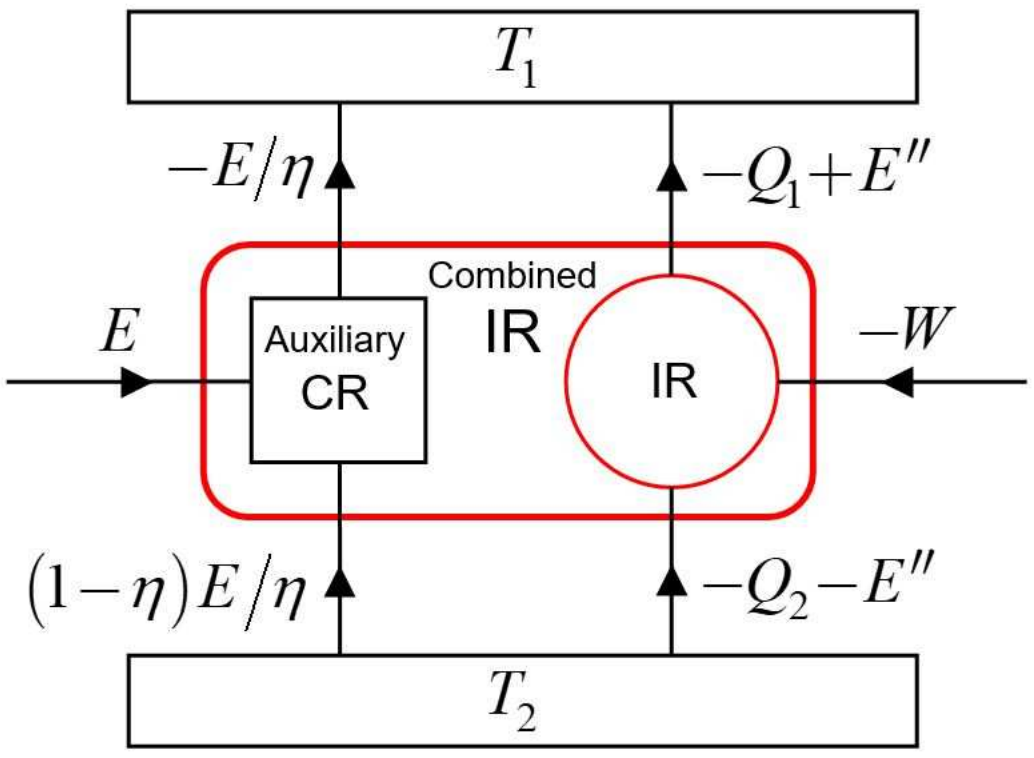

**Figure 5.** IR of Fig. 3(b) together with an auxiliary CR adjusted to receive the work *E*. As no extra irreversibility is added by the CR, by imposing (9), the combined IR thus obtained is *indistinguishable* from that of Fig. 2(b).

In contrast to *entropy generation* $\Delta S_G$, the total *lost work* $W_L$ cannot be determined from Figs. 2 and 3, because these diagrams do not give information about what exactly is going on inside the system that performs the cycle. Until the system is examined in more detail, what can be said about the lost work $W_L$ in an irreversible cycle is restricted to (3), and it is not possible to derive an explicit relation between $W_L$ and any of the parameters $E$, $E'$ or $E''$.

In the next section, two irreversible engines are analyzed, and it is highlighted that the entropy generation can be determined by lost work and system temperatures, as an alternative to the usual procedure that uses the heat and reservoir temperatures.

## 3. Two illustrative examples

We will consider two quasistatic *irreversible* cycles that exchange heat with only two reservoirs, operating as heat engines. Each of the irreversible cycles corresponds to the diagram of Fig. 2(a) and will be compared to the CE of Fig. 1(a), whose $Q_1$, $Q_2$ and *W* of will be taken as reference values. The goal is to show that the entropy generation $\Delta S_G$ can be determined either by (2) or (3), i.e. by using heat and reservoir temperatures, which is the standard procedure, or by using lost work and system temperatures, which is a new approach.

We will assume that the system performing the cycle is an ideal gas, denoting the amount of gas in moles by *n*, the universal gas constant by *R*, and the thermal capacity at constant volume by $C_V$. So, we have two important relations

$$PV = nRT \tag{11}$$

and

$$\mathrm{d}U = C_V \mathrm{d}T, \tag{12}$$

where *P*, *V* and *U* stand for pressure, volume, and internal energy, respectively.

## 3.1 Irreversible Stirling engine

The first heat engine we consider is the irreversible Stirling engine, operating in the clockwise direction 12341, as shown in Fig. 6. This cycle has two isothermal processes and two isochoric processes and was studied in Ref. [1]. The cycle $1'2'3'41'$ is the CE that absorbs the same heat $Q_1$ from the reservoir at temperature $T_1$, which is shown for comparison purposes.

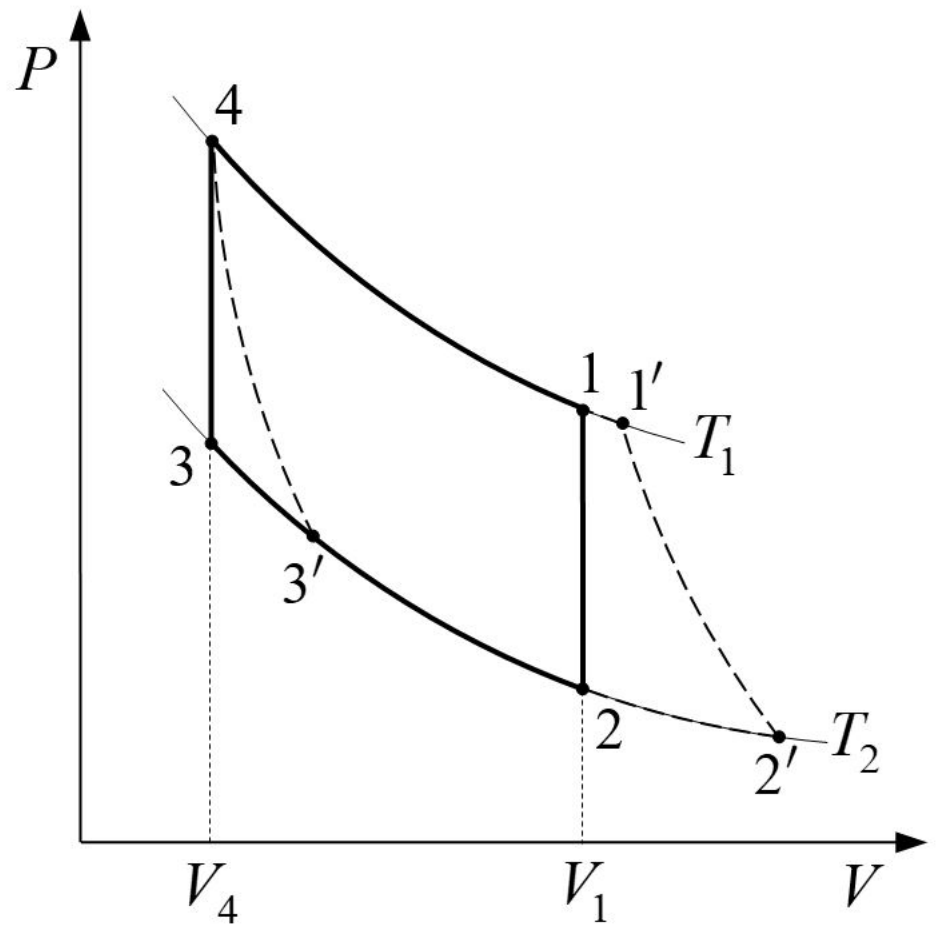

**Figure 6.** Pressure-volume diagram of the irreversible Stirling cycle 12341. The CE that absorbs the same heat $Q_1$ at reservoir temperature $T_1$ is the cycle $1'2'3'41'$.

In processes 12 and 23, the system delivers heat to the reservoir at temperature $T_2$, and in processes 34 and 41, the system receives heat from the reservoir at temperature $T_1$, with $T_1 > T_2$.

The entropy generation $\Delta S_G$ was determined in Ref. [1] using (2), but here we will use (3) instead, i.e. we will use the lost work and system temperature, which is a new and instructive approach. The lost work is defined, as we have already said, by $\delta W_L = T dS_G$ and can be written as [4] (see Appendix)

$$\delta W_L = \left(1 - \frac{T}{T_e}\right) \delta Q + \delta W_D, \tag{13}$$

where $\delta W_D$ is the dissipative work given by

$$\delta W_D = P dV + P_e dV_e. \tag{14}$$

The first term in (13) is the lost work due to heat exchange under a finite temperature difference, while the second term is the lost work due to dissipation of mechanical energy, e.g. by friction. In this example, it is assumed that $\delta W_D = 0$. Thus, by inserting (13) into (3), we get

$$\Delta S_G = \oint \frac{\delta W_L}{T} = \oint \left(\frac{1}{T} - \frac{1}{T_e}\right) \delta Q. \tag{15}$$

Integration is carried out only in processes where entropy generation exists, i.e. in isochoric processes $12$ and $34$. For these processes, $\delta W = 0$, and, by the first law, $\delta Q = \mathrm{d}U$. By (12), we have $\delta Q = C_V\,\mathrm{d}T$ and (15) becomes

$$\Delta S_\mathrm{G} = C_V \left( \int_{T_1}^{T_2} \frac{\mathrm{d}T}{T} - \int_{T_1}^{T_2} \frac{\mathrm{d}T}{T_2} + \int_{T_2}^{T_1} \frac{\mathrm{d}T}{T} - \int_{T_2}^{T_1} \frac{\mathrm{d}T}{T_1} \right), \tag{16}$$

which gives

$$\Delta S_\mathrm{G} = C_V \frac{\left(T_1 - T_2\right)^2}{T_1 T_2} > 0 . \tag{17}$$

This result, obtained using lost work $\delta W_\mathrm{L}$ and system temperature *T*, is the same as that obtained in Ref. [1] by using heat $\delta Q$ and reservoir temperatures $T_\mathrm{e}$, as it should. This reveals the fundamental role of $\delta W_\mathrm{L}$ and its close relation with $\mathrm{d}S_\mathrm{G}$. The total lost work $W_\mathrm{L}$ is given by

$$W_\mathrm{L} = \oint \delta W_\mathrm{L} = \oint \left( 1 - \frac{T}{T_\mathrm{e}} \right) C_V\,\mathrm{d}T , \tag{18}$$

$$W_\mathrm{L} = C_V \left( \int_{T_1}^{T_2} \mathrm{d}T - \int_{T_1}^{T_2} \frac{T\mathrm{d}T}{T_2} + \int_{T_2}^{T_1} \mathrm{d}T - \int_{T_2}^{T_1} \frac{T\mathrm{d}T}{T_1} \right), \tag{19}$$

which gives

$$W_\mathrm{L} = C_V \frac{\left(T_1 - T_2\right)^2 \left(T_1 + T_2\right)}{2 T_1 T_2} > 0 . \tag{20}$$

Comparing (17) with (20), we get for this cycle

$$W_\mathrm{L} = \frac{T_1 + T_2}{2} \Delta S_\mathrm{G} . \tag{21}$$

The heats and work, $Q_1$, $Q_2 - E$ and $W_\mathrm{I}$, are given by

$$Q_1 = \int_3^4 \delta Q + \int_4^1 \delta Q = \int_{T_2}^{T_1} C_V \mathrm{d}T + \int_{V_4}^{V_1} P\mathrm{d}V , \tag{22}$$

$$Q_1 = C_V \left(T_1 - T_2\right) + nRT_1 \ln\frac{V_1}{V_4} > 0 ; \tag{23}$$

$$Q_2 - E = \int_1^2 \delta Q + \int_2^3 \delta Q = \int_{T_1}^{T_2} C_V \mathrm{d}T + \int_{V_2}^{V_3} P\mathrm{d}V , \tag{24}$$

$$Q_2 - E = -C_V \left(T_1 - T_2\right) + nRT_2 \ln\frac{V_3}{V_2} < 0 ; \tag{25}$$

$$W_\mathrm{I} = W + E = -Q_1 - \left(Q_2 - E\right) = -nR\left(T_1 - T_2\right) \ln\frac{V_1}{V_4} < 0 , \tag{26}$$

where we have used the equality $\ln\left(V_2/V_3\right) = \ln\left(V_1/V_4\right)$.

By (23), the work $W$ for the CE $1'2'3'41'$, which absorbs the same $Q_1$, is

$$W = -\left(1 - \frac{T_2}{T_1}\right) Q_1 = -C_V \frac{(T_1 - T_2)^2}{T_1} - nR(T_1 - T_2) \ln \frac{V_1}{V_4} < 0. \tag{27}$$

Thus, the excess heat rejected $E = W_I - W$ becomes

$$E = W_I - W = C_V \frac{(T_1 - T_2)^2}{T_1} > 0, \tag{28}$$

which is different from $W_L$ given by (20). Because cycle 12341 and the CE $1'2'3'41'$ absorb the same heat $Q_1$, (6) holds. This can be verified by dividing (28) by $T_2$, which gives (17).

### 3.2 Irreversible engine with two adiabatics and two isotherms

The second example is the cycle 12341 shown in Fig. 7. It operates in the clockwise direction, i.e. as a thermal engine, and has two isothermal processes and two adiabatic processes. It looks like a CE, but there is dissipative work (e.g. friction) during the expansion corresponding to the isothermal process at temperature $T_1$, so the cycle is irreversible. The CE that absorbs the same heat $Q_1$ is the cycle $1'2'341'$, which is shown for comparison purposes.

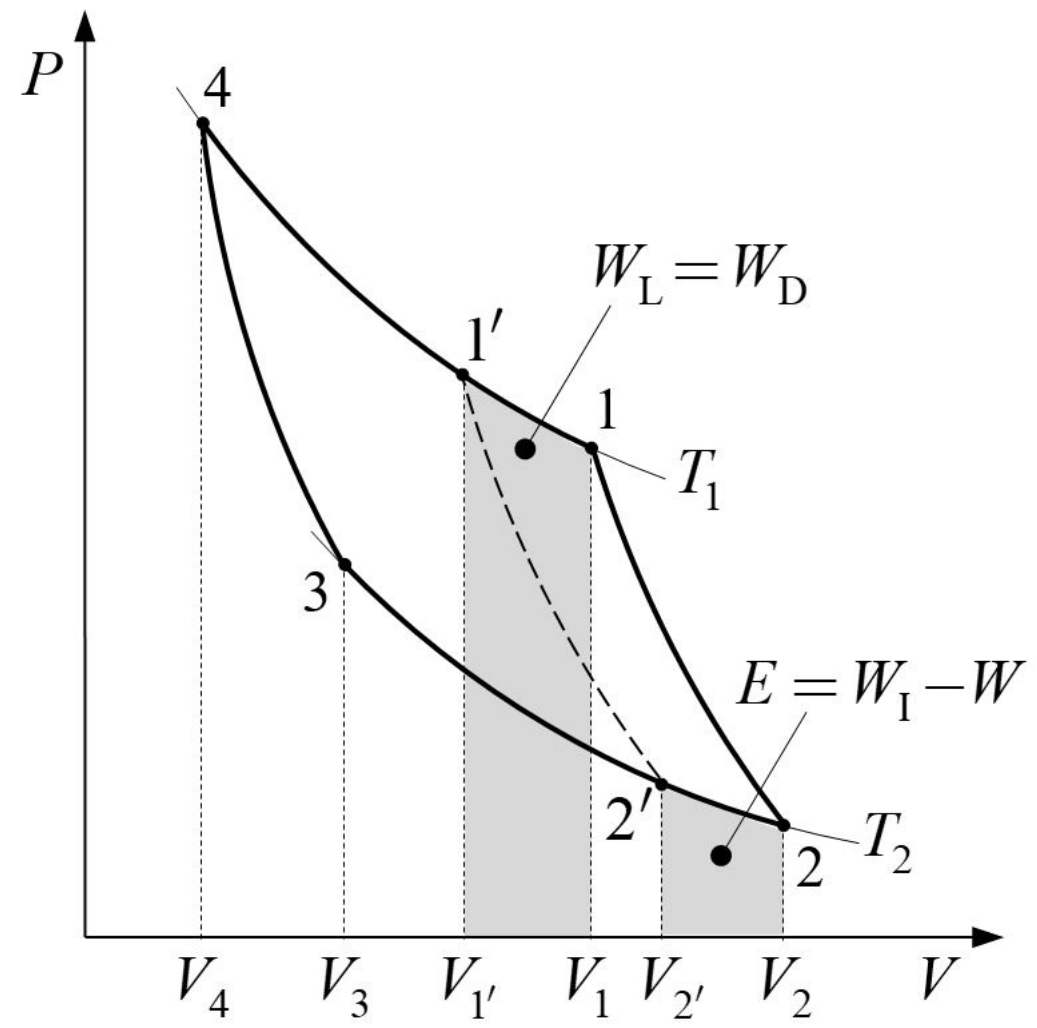

**Figure 7.** Pressure-volume diagram of a heat engine 12341, having two adiabatics and two isotherms. In the expansion 41, corresponding to the isotherm at temperature $T_1$, there is dissipative work $W_D$ (e.g. friction) so that the cycle is irreversible. The CE $1'2'341'$ absorbs the same heat $Q_1$ from the reservoir at temperature $T_1$.

By (A1), (A7) and (A9) of the Appendix, in expansion 41 the area under it to the volume axis is equal to minus the work $-W_{41}$ plus the dissipative work $W_D$, i.e.

$$\int_4^1 P \mathrm{d}V = -W_{41} + W_D, \tag{29}$$

where

$$-W_{41} = \int_4^{1'} P\mathrm{d}V = nRT_1 \ln \frac{V_{1'}}{V_4}, \tag{30}$$

$$W_\mathrm{D} = \int_{1'}^{1} P\mathrm{d}V = nRT_1 \ln \frac{V_1}{V_{1'}}. \tag{31}$$

Please note that in a quasistatic irreversible process where $\delta W_\mathrm{D} \neq 0$ the area under the curve in a *P-V* diagram is not equal to the value of *W*, as is thoroughly explained in Ref. [11].

Unlike the previous example, the lost work here is just dissipative work $\delta W_\mathrm{D}$, i.e. (13) becomes $\delta W_\mathrm{L} = \delta W_\mathrm{D}$. Since $\delta W_\mathrm{D}$ takes place at constant system temperature $T_1$, by (3) and (31), the entropy generation is

$$\Delta S_\mathrm{G} = \oint \frac{\delta W_\mathrm{L}}{T} = \frac{W_\mathrm{D}}{T_1} = nR \ln \frac{V_1}{V_{1'}} > 0. \tag{32}$$

By (12), the heat exchanged with the reservoirs is equal to minus the work in the respective isotherms. So, using (30), we have

$$Q_1 = -W_{41} = nRT_1 \ln \frac{V_{1'}}{V_4} > 0, \tag{33}$$

$$Q_2 - E = -W_{23} = nRT_2 \ln \frac{V_3}{V_2} < 0. \tag{34}$$

With the above $Q_1$ and $Q_2 - E$ we can use (2) to re-obtain (32),

$$\Delta S_\mathrm{G} = \oint \frac{-\delta Q}{T_\mathrm{e}} = \frac{-Q_1}{T_1} + \frac{-Q_2 + E}{T_2} = nR \ln \frac{V_1}{V_{1'}} > 0, \tag{35}$$

where we used the equality $\ln(V_2/V_3) = \ln(V_1/V_4)$.

The irreversible work $W_\mathrm{I}$ is given by

$$W_\mathrm{I} = -Q_1 - (Q_2 - E) = -nRT_1 \ln \frac{V_{1'}}{V_4} - nRT_2 \ln \frac{V_3}{V_2} < 0, \tag{36}$$

and the work *W*, for the CE $1'2'341'$ exchanging the same heat $Q_1$, is

$$W = -nRT_1 \ln \frac{V_{1'}}{V_4} - nRT_2 \ln \frac{V_3}{V_{2'}} < 0. \tag{37}$$

Thus, the excess heat rejected $E = W_\mathrm{I} - W$ becomes

$$E = W_\mathrm{I} - W = nRT_2 \ln \frac{V_2}{V_{2'}} > 0, \tag{38}$$

which is different from $W_\mathrm{L}$ given by (31). Because cycle 12341 and the CE $1'2'341'$ absorb the same heat $Q_1$, (6) holds. Indeed, as $\ln(V_2/V_{2'}) = \ln(V_1/V_{1'})$, dividing (38) by $T_2$ we get (35).

## 4. Conclusions

We have shown that the *lost work* cannot be determined from the usual diagrams for irreversible engine and irreversible refrigerator. Using the concept of identical processes, we have highlighted that what characterizes the irreversibility is the entropy generation, which can be determined either in terms of heat and reservoir temperatures or in terms of lost work and system temperatures: the two sides of the same coin. The definition of lost work as stated in this paper has a clear relation with entropy generation and is universal, thus not depending on any assumptions beyond considering the surroundings consisting of reservoirs. It applies to all processes, cyclic or otherwise. Of course, the difference between irreversible and reversible work may be useful in some contexts, but it has no definite relation with entropy generation because it depends on the problem and the conditions imposed. Two illustrative irreversible cycles have been analyzed, which have provided a deeper insight into the role of lost work in entropy generation. Finally, it is worth noting that the discussion carried out in this paper is new and emphasizes the strength of the *concept of identical processes* in clarifying some scientific and educational issues regarding irreversible cycles.

## Appendix

Everything that follows is covered in full and in more detail in Ref. [4], but for the sake of completeness, what is needed for this paper is briefly presented here.

The equation that contains *all* the thermodynamic information about a process is given by

$$T\mathrm{d}S - P\mathrm{d}V = -T_\mathrm{e}\mathrm{d}S_\mathrm{e} + P_\mathrm{e}\mathrm{d}V_\mathrm{e}, \qquad \text{(A1)}$$

where $T$, $S$, $P$ and $V$ denote temperature, entropy, pressure, and volume, respectively, and the subscript ‘e’ stands for surroundings variables. The entropy variations in the system $\mathrm{d}S$ and surroundings $\mathrm{d}S_\mathrm{e}$ are related to the *entropy flow* $\mathrm{d}S_\phi$ and the *entropy generation* $\mathrm{d}S_\mathrm{G}$ by

$$\mathrm{d}S = \mathrm{d}S_\phi + \beta\,\mathrm{d}S_\mathrm{G}, \qquad \text{(A2)}$$

$$\mathrm{d}S_\mathrm{e} = -\mathrm{d}S_\phi + \beta_\mathrm{e}\mathrm{d}S_\mathrm{G}, \qquad \text{(A3)}$$

where $\beta$ and $\beta_\mathrm{e}$ are arbitrary non-negative parameters satisfying

$$\beta + \beta_\mathrm{e} = 1. \qquad \text{(A4)}$$

By making $\beta = 1$ and $\beta_\mathrm{e} = 0$ in (A2) and (A3) (i.e. *considering that surroundings consist of reservoirs*), inserting these equations into (A1) and using the following definitions

$$\delta W_\mathrm{L} = T\mathrm{d}S_\mathrm{G} \text{ (lost work)}, \qquad \text{(A5)}$$

$$\delta Q = -T_\mathrm{e}\mathrm{d}S_\mathrm{e} \text{ (heat)}, \qquad \text{(A6)}$$

$$\delta W_\mathrm{D} = P\mathrm{d}V + P_\mathrm{e}\mathrm{d}V_\mathrm{e} \text{ (dissipative work)}, \qquad \text{(A7)}$$

the *lost work* $\delta W_{\mathrm{L}}$ can be written as

$$\delta W_{\mathrm{L}} = \left(1 - \frac{T}{T_{\mathrm{e}}}\right) \delta Q + \delta W_{\mathrm{D}} . \tag{A8}$$

The dissipative work $\delta W_{\mathrm{D}}$ given by (A7) is the difference between *work*

$$\delta W = P_{\mathrm{e}} \mathrm{d} V_{\mathrm{e}} \tag{A9}$$

and *configuration work*

$$\delta W_{\mathrm{C}} = -P \mathrm{d} V . \tag{A10}$$

For a finite process, from (A5) and (A6) we have

$$\int \frac{\delta W_{\mathrm{L}}}{T} = \Delta S_{\mathrm{G}} , \tag{A11}$$

$$\int \frac{-\delta Q}{T_{\mathrm{e}}} = \Delta S_{\mathrm{e}} . \tag{A12}$$

Finally, for cyclic processes we have $\Delta S = 0$ and thus $\Delta S_{\mathrm{e}} = \Delta S_{\mathrm{G}}$, both (A11) and (A12) then become alternatives for obtaining the entropy generation,

$$\oint \frac{\delta W_{\mathrm{L}}}{T} = \Delta S_{\mathrm{G}} , \tag{A13}$$

$$\oint \frac{-\delta Q}{T_{\mathrm{e}}} = \Delta S_{\mathrm{G}} . \tag{A14}$$